\documentclass[12pt]{article}
\usepackage{latexsym,graphicx,multirow}
\usepackage{float}
\usepackage{amssymb}
\usepackage{amsmath}
\usepackage{amscd}
\usepackage{amsthm}
\usepackage[left=2cm,top=2.5cm,right=2.5cm,bottom=1.5cm]{geometry}
\usepackage{hyperref}
\usepackage{epstopdf}
\usepackage{cite}

\begin{document}
	
\begin{center}
\large{\bf{A new class of holographic dark energy models in LRS Bianchi Type-I }} \\
		\vspace{10mm}
		\normalsize{ Anirudh Pradhan$^1$, Vinod Kumar Bhardwaj$^2$, Archana Dixit$^3$, Syamala Krishnannair$^{4}$  }\\
		\vspace{5mm}
		\normalsize{$^{1,2,3}$Department of Mathematics, Institute of Applied Sciences and Humanities, GLA University,
			Mathura-281 406, Uttar Pradesh, India}\\
		\vspace{5mm}
		\normalsize{$^{4}$Department of Mathematical Sciences, Faculty of Science, Agriculture and Engineering,University of 
		Zululand, Kwadlangezwa 3886, South Africa} \\
		
		\vspace{2mm}
		$^1$E-mail: pradhan.anirudh@gmail.com \\
		\vspace{2mm}
         $^2$E-mail:dr.vinodbhardwaj@gmail.com\\
         \vspace{2mm}
         $^3$E-mail:archana.dixit@gla.ac.in\\
         \vspace{2mm}
         $^{4}$E-mail:krishnannairs@unizulu.ac.za\\
	\vspace{10mm}
		
		%\date{}
		%\maketitle
		
	\end{center}
\begin{abstract}
In this paper, we examine the (LRS) Bianchi-type-I cosmological model with holographic dark energy. The exact solutions to the corresponding field 
equations are obtained  by using generalized hybrid expansion law (HEL). The EoS parameter $\omega$ for DE  is found to be time dependent and 
redshift dependent and its exiting range for derived model is  agreeing well with the current observations. Here we likewise apply two mathematical 
diagnostics, the statefinder ({r, s}) and $\omega_{d}-\omega^{'}_{d}$ plane to segregate HDE model from the $\Lambda CDM$ model. 
Here the $\omega_{d}-\omega^{'}_{d}$ diagnostic trajectories is the good tool to classifying the dynamical DE model. We found that our 
model lies in both thawing region  and freezing region. We have also construct the potential as well as dynamics of the quintessence and 
tachyon scalar field. Some physical and geometric properties of this model along with the physical acceptability of cosmological solution 
have been discussed in detail.
\end{abstract}
	
	
%%%%%%%%%%%%%%%%%%%%%%%%%%%%%%%%%%%%%%%%%%%%%%%%% Section 1 %%%%%%%%%%%%%%%%%%%%%%%%%%%%%%%%%%%%%%%%%%%%%%%%%%%%%%%%%%%%%%%%%%%%%%
\section{Introduction}
Motivated by the discovery of ``black hole thermodynamics \cite{ref1,ref2} Hooft proposed the well known holographic principle (HP)" \cite{ref3}. 
As a advanced form of ``Plato's cave ", the HP describes that all of the info confined in a bulk of space may be visualized as a hologram, 
which correlates to a concept positioning on the frontier of that space. Afterward, Susskind \cite{ref4} suggest the exact string hypothesis 
thought of the rule. Besides, Maldacena \cite{ref5} recommended the notable `` AdS/CFT correspondence" in 1997, which is the best acknowledgment 
of Holographic Principle. Presently it notable that Holographic Principle is normally used to be an essential idea of quantum gravity.\\ 

According to the basic concepts of quantum gravitational, we can also investigate the nature of DE, by HDE principle. As per this rule, 
the `degrees of freedom' in a constrained structure must be finite and not measure with the volume of the system \cite{ref6}.
As we mention above that HP is the essential part of ``quantum gravity" and has significant potential to solve several long-lasting problems 
in different fields. The HP can also be used to solve the DE problem. The HP reveal us that all the physical measures of the universe, 
like energy density of DE ($\rho_{d}$), can be depicted by certain amounts on the limit of the universe. This indicates that only two physical 
measures ``the diminished Planck mass $M_{p}$ and the cosmological length scale $L$", can be utilized to develop the expression of $\rho_{d}$. 
In view of the dimensional analysis, we have
\begin{equation}\label{1}
\rho_{d} =c _{1} M_{p}^{4}+c_{2} M_{p}^{4} L^{-2}+c_{3} L^{4-}+........
\end{equation}
here $c_{1}, c_{2}, c_{3}$ are constants. Author \cite{ref7} suggest that the $c_{1}$ expression is not viable with Holographic principle. 
According to the HP, the local quantum field (LQF) theory should not be a decent characterization for a black hole. Particularly, the traditional 
estimate $\rho_{d}\approx c_{1} M^{4}_{P}$ the (LQF) this explanation should not include any theory. The LQF theory attains a non-trivial 
`UV cutoff’ $\Lambda$. As a consequence of the QF theory, the vacuum fluctuation calculated is $\rho_{d}\approx c_{1} M_{4}^{p} L^{-2}$. 
The suggestion of HDE was depend on the concept of (QF) theory, that a short distance cut-off is associated to a long-distance cut-off due 
to the limit set by the establishment of a black hole \cite{ref8}-\cite{ref14}.  \\

We compared only second term and neglecting other terms then we get HDE as
\begin{equation}\label{2}
\rho_{d} = 3C^{2}M^{2}_{p}L^{-2}
\end{equation}
where $C$ is other constant. In Eq.(\ref{1}) $c_{1}$ term is not present in this expression. It is important to note that this expression 
of $\rho_{d}$ is obtained by dimensional analysis and applying HP, 
rather than adding a DE term to the Lagrangian. Because of this remarkable feature HDE theory distinguished from other DE theories. In the 
cosmological context Fischler and Susskind \cite{ref15} suggested a new version of the holographic concept in the cosmological evolution. 
In the framework of the DE problem, HP provides a relation $\rho =H^{2}$. Granda and Olivers suggested a holographic density of 
the type $\rho\approx\alpha H^{2} +\beta \dot H$ in \cite{ref16}, where $\alpha$ and $\beta$ are constants. They contend that their new DE model 
describes the universe's  accelerated expansion and it is consistent with current observational findings. Bianchi-type models are the 
most basic and accurate anisotropic models, completely describing anisotropic effects.
Chimento {\it et al.} \cite{ref16a} looked at a Universe with an interacting dark sector and a decoupling component that may be used to simulate a radiation term. 
	\cite{ref16b} discusses a model of interacting dark matter and dark energy based on a modified holographic Ricci cutoff in a spatially flat FRW spacetime. In \cite{ref16c}, a universe with interacting dark matter, holographic dark energy, and dark radiation is examined for FRW spacetime. FRW universe with interacting dark matter, modified holographic Ricci dark energy, and a decoupled baryonic component was studied by Chimento et al. \cite{ref16d}. Chimento {\it et al.} \cite{ref16e} also introduced scalar field models through internal space structure generalization of the quintom cosmological scenario. \\

In this manner numerous cosmologist have proposed the various aspects of these spatially homogeneous and anisotropic Bianchi-type models in various 
modified theories \cite{ref17}-\cite{ref24}. Zadeh \cite{ref25} examined the astronomical advancement of THDE in Bianchi type-I model 
with various IR cut-offs. Author \cite{ref26,ref27,ref28} dealt with different anisotropic DE cosmological models . 
Authors \cite{ref29,ref30}  worked on  the general class of anisotropic cosmological models in particular theories.\\

In the literature, anisotropic Bianchi type-I, III, and V DE models with the standard perfect fluid have  been also widely researched 
\cite{ref31}-\cite{ref37}. Sarkar and Mahanta \cite{ref38} developed a HDE cosmological model in Bianchi type-I space time with quintessence. 
Adhav et al. \cite{ref39} investigated an  homogeneous and anisotropic Bianchi type-I universe field with interacting dark matter and HDE. 
A similar way Pradhan and Amirhashchi \cite{ref40} constructed a accelerating DE model and investigated some new precise solutions of 
Einstein's field conditions in a spatially and anisotropic Bianchi type-V space-time.\\

The Bianchi type-I string cosmological model with bulk viscosity was examined by Tiwari and Sonia \cite{ref41} . Many cosmologists 
\cite{ref42,ref43,ref44} have explored the Bianchi type-I models from a different perspective. Now we can say that Spatially homogeneous and 
anisotropic cosmological models play an important role in the description of large-scale phenomena in the cosmos. Among the numerous dynamical 
DE models, the cosmological constant is the most important component but it suffers from `cosmic coincidence and fine tuning' problems. 
As a result of this explanation, various dynamical DE models have been recommended and they were given as
family of `scalar field models ( quintessence, k-essence,  phantom, ghost,
etc.), Chaplygin gas as well as holographic DE (HDE)' models. The authors \cite{ref45} concluded that the modified structure scalars play 
an important role in the dynamics of compact object. The authors \cite{ref46} investigated the oscillatory behaviour of the anisotropy 
Bianchi -I spacetime. and emphasized the presence of a residue in the energy density in a late time isotropic universe coming 
from its past anisotropic behaviour. Renyi HDE correspondence is discussed in higher dimensional Kaluza-Klein cosmology \cite{ref47}. \\

In this manuscript  we have chosen LRS Bianchi types-I (LRSBT-I) homogeneous and isotropic model. This model is very important to study space-times 
where anisotropy occurs at early stage and isotropy at later stage of the evolution of the universe. We have developed various cosmological 
parameters to describe the dynamics of universe. These parameters are evaluated in terms of $t$ and redshift $z$. Our work is outlined in the 
following manner: In  Sec. $2$, we formulate the basic field equations for LRSBT-I metric. The solutions and the physical behaviour of the 
cosmological parameters are discussed in Sec. $3$. In Sect. $4$, we have discussed the state finder diagnosis. In Sec. $5$, we have studied 
the distinct regions in $\omega_{d}- \omega_{d}^{'}$ plane. In Sec. $6$, we have constructed the DE scalar field models for tachyon and quintessence. 
The nature of the deceleration parameter ($q$) is discussed in Sec. $7$. We have concluded the results in the last Sec. $8$.

%%%%%%%%%%%%%%%%%%%%%%%%%%%%%%%%%%%%%%%%%%%%%%%%%%%%% Section 2 %%%%%%%%%%%%%%%%%%%%%%%%%%%%%%%%%%%%%%%%%%%%%%%%%%%%%%%%%%%%%%%%%%%%

\section{ LRS Bianchi type-I universe and Basic Field Equations}

In this cosmological model we consider the LRS Bianchi type I metric of the form
\begin{equation}\label{3}
ds^2=-dt^2+A(t)^2 dx^2+B(t)^2 \left( dy^2+ dz^2 \right),
\end{equation}
where the metric coefficients ``$ A$ and $ B $" are function of $t$ only. \\

In general relativity, Einstein's field equation with the cosmological constant:

\begin{equation}\label{4}
R_{ij}-\frac{1}{2} g_{ij} R+\Lambda g_{ij} =-\left(T_{ij}+\bar{T}_{ij}\right)
\end{equation}

The notations have their typical significance. The new HDE for actual interpretation and the energy momentum tensor for matter can be composed as
: $ T_{ij} = \rho_{m} u_{i} u_{j}  $, $ \bar{T}_{ij}=(\rho_{d} +p_{d}) u_{i} u_{j} + g_{ij} p_{d} $, where $\rho_{m}$ and $\rho_{d}$ address 
the matter energy density and new HDE density, and $ p_{d} $ stands for the HDE pressure. The field equations given by Eq. (\ref{4}) along with 
the energy momentum tensors defined above for the LRSBT-I metric of three independent equations:
\begin{equation}
\label{5}
2 \frac{\ddot{B}}{B}+ \frac{\dot{B}^2}{B^2}=-p_{d}+\Lambda,
\end{equation}
\begin{equation}\label{6}
\frac{\ddot{A}}{A}+\frac{\ddot{B}}{B}+ \frac{\dot{A} \dot{B} }{A B}=-p_{d}+\Lambda,
\end{equation}
\begin{equation}\label{7}
2\frac{\dot{A} \dot{B}}{AB}+\frac{\dot{B}^2}{B^2}=\rho_{m}+\rho_{d}+\Lambda .
\end{equation}

Now, as indicated in Refs. \cite{ref48,ref49}, the HDE density with IR cut-off is defined as:

\begin{equation}\label{8}
\rho_{d}=3 M^{2}_{p} (k H^2+m \dot{H}),
\end{equation}
where $ k $ and $ m $ are the dimensionless parameters and must satisfy the restrictions imposed by the current observational framework 
with $  M^{2}_{p} = 8\pi G = 1 $. The energy conservation law is determined as $ T^{ij}_{;j} = 0 $, gives 
$ \rho_{m} +\dot{\rho_{d}} + 3H(\rho_{m} + \rho_{d} + p_{d}) = 0 $. The energy conservation law for matter and 
dark energy are, separately expressed as $ \dot{\rho_{m}} +3H \rho_{m} = 0 $ and $ \dot{\rho_{d}} +3H(\rho_{d} +p_{d}) = 0 $ respectively. Here, $H$ is the Hubble parameter defined by $ H=\frac{\dot{a(t)}}{a(t)}$, where the dot represents a time derivative. The Hubble parameter varies with time, not with space, being the Hubble constant $H_{0}$ is the current value. The relative expansion of the universe is parametrized by a dimensionless scale factor $a$. The Hubble constant ($H_{0}$) is important as it is needed to estimate the size and the age of the Universe. $H$  indicates the rate at which the universe is expanding, from the pre mordial ``Big Bang". \\

%%%%%%%%%%%%%%%%%%%%%%%%%%%%%%%%%%%%%%%%%%%%%%%%%%%%%% Section - 3 %%%%%%%%%%%%%%%%%%%%%%%%%%%%%%%%%%%%%%%%%%%%%%%%%%%%%%%%%%%%%%%%%%%%%%%%

\section{Solutions and Physical Behaviour of the Model}

Here, we have three equations  Eq. (\ref{5})- Eq. (\ref{7})  with five unknown parameters namely A, B, $\rho_{d}$, $\rho_{m}$ and $\Lambda$. 
For solving these equations,  we need two more physical assumption. Our first assumption has to taken in  Eq. (\ref{8}) and the second assumption 
is the scale factor $a(t)$ which  is defined as \cite{ref50,ref51} :
\begin{equation}\label{9}
	a=t^{\alpha} e^{\beta t}
\end{equation}
here $\alpha$ and $\beta$ are constant. The scale factor is the fundamental function that characterises the global features of our Universe. For the analysis of various problems, an exact equation for this scale factor as a function of physical and conformal time is desirable \cite{ref51a}. In cosmology, the scale factor is a parameter that represents how the universe's size is changing in relation to its current size.\\ 

This results  have a linear relationship between the directional Hubble parameters as $ H_{1}=n H_{2} $ where $n$ is positive constant. 
Here $A= \left(t^{\alpha} e^{\beta t}\right)^{\frac{3 n}{(n+2)}}$, $B=\left(t^{\alpha} e^{\beta t}\right)^{\frac{3}{(n+2)}}$, then we obtained
$H_{1}=\frac{\dot A}{A}=\frac{3 n}{(n+2)}\left(\frac{\alpha}{t}+\beta\right)$, $ 
H_{2}=\frac{\dot B}{B}=\frac{3 }{(n+2)}\left(\frac{\alpha}{t}+\beta\right)$ are the directional Hubble Parameters. The Hubble Parameter is derived as:

\begin{equation}\label{10}
	H=\left(\frac{\alpha}{t}+\beta\right)
\end{equation}
By using  Eq. (\ref{9}) we obtained deceleration parameter $q= \frac{\alpha}{(\alpha+\beta t)^{2}}-1$. This defines the relation  
$\beta = \dfrac{1}{13.8}[\left(\frac{\alpha}{0.27}\right)^{1/2}-\alpha]$, between $\alpha$ and $\beta$, by using 
$t_{0}=13.8 Gry$  $q_{0} =-0.73$, we get the corresponding value of $\beta$ \cite{ref52,ref53} [See Table - 1]. 
From this table, we observe that $\alpha$ increases with the increase of $\beta$. For $0 < \alpha \leq 1$, $\beta$ increases  
but for $\alpha > 1$, $\beta$ starts decreasing. In all figures we have considered the three sets of ($\alpha$, $\beta$) as 
($0.25$, $0.0516$), ($0.50$, $0.0624$) and ($0.75$, $0.0665$) respectively. It is worth mentioned that for $0<\alpha<1$, our models 
show the transitioning scenario whereas for $\alpha \geq 1$, the models are in accelerating phase only.  \\

Using Eq. (\ref{9}) in (\ref{8}), we obtain the HDE of the model as:

\begin{equation}\label{11}
\rho_{d}=\frac{3 \left(k (\alpha +\beta  t)^2-\alpha  m\right)}{t^2}.
\end{equation}

By using Eqs. (\ref{10})-(\ref{11}) in energy conservation law $ \dot{\rho_{d}} +3H(1+\omega_{d})\rho_{d} = 0 $, we get the EoS parameter for the model as:

\begin{equation}\label{12}
\omega_{d}=-\frac{2 \alpha  (m-k (\alpha +\beta  t))}{3 (\alpha +\beta  t) \left(k (\alpha +\beta  t)^2-\alpha  m\right)}-1.
\end{equation}
Solving Eqs. (\ref{5})-(\ref{7}) and using Eqs. (\ref{9})-(\ref{11}), the expressions for different cosmological parameters, $p_{d}$, $\rho_{m}$ and 
$\Lambda$ are obtained as
\begin{equation}\label{13}
p_{d}=\frac{3 \left(k (\alpha +\beta  t)^2-\alpha  m\right)}{t^2} \left(-\frac{2 \alpha  (m-k (\alpha +\beta  t))}{3 (\alpha +\beta  t) 
\left(k (\alpha +\beta  t)^2-\alpha  m\right)}-1\right),
\end{equation}
\begin{equation}\label{14}
\rho_{m}=\frac{2}{t^2} \left(\frac{\alpha  (-k) (n+2)^2+9 \alpha ^2 (n-1)+9 \beta ^2 (n-1) t^2+3 \alpha  (6 \beta  n t+n-6 \beta t+2)}{(n+2)^2}+
\frac{\alpha  m}{\alpha +\beta  t}\right),
\end{equation}

\begin{equation}\label{15}
\Lambda=\frac{1}{t^2} \left(2 \alpha  k+3 \left(\alpha  m-k (\alpha +\beta  t)^2\right)-\frac{2 \alpha  m}{\alpha +\beta  t}-
\frac{6 \alpha }{n+2}+\frac{27 (\alpha +\beta  t)^2}{(n+2)^2}\right).
\end{equation}
	
The matter density parameter $\Omega_{m}$ and HDE density parameter $\Omega_{d}$
are given by $ \Omega_{m}=\frac{\rho_{m}}{3 H^2} $ and $ \Omega_{d}=\frac{\rho_{d}}{3 H^2}$.
\begin{eqnarray}\label{16}
\Omega_{m}=\frac{\alpha  (-k) (n+2)^2+18 \alpha ^2 (n-1)+18 \beta ^2 (n-1) t^2+6 \alpha  (6 \beta  n t+n-6 \beta  t+2)}
{ 3 (2\alpha +\beta  t)^2 (n+2)^2}+\frac{2 \alpha  m}{3 (\alpha +\beta  t)^3}
\end{eqnarray}
\begin{equation}\label{17}
\Omega_{d}=k-\frac{\alpha  m}{(\alpha +\beta  t)^2}
\end{equation}
 
 %%%%%%%%%%%%%%%%%%%%%%%%%%%%%%%%%% Figure 1 %%%%%%%%%%%%%%%%%%%%%%%%%%%%%%%%%%%
 	\begin{figure}[H]
	\centering
	(a)\includegraphics[width=7cm,height=6cm,angle=0]{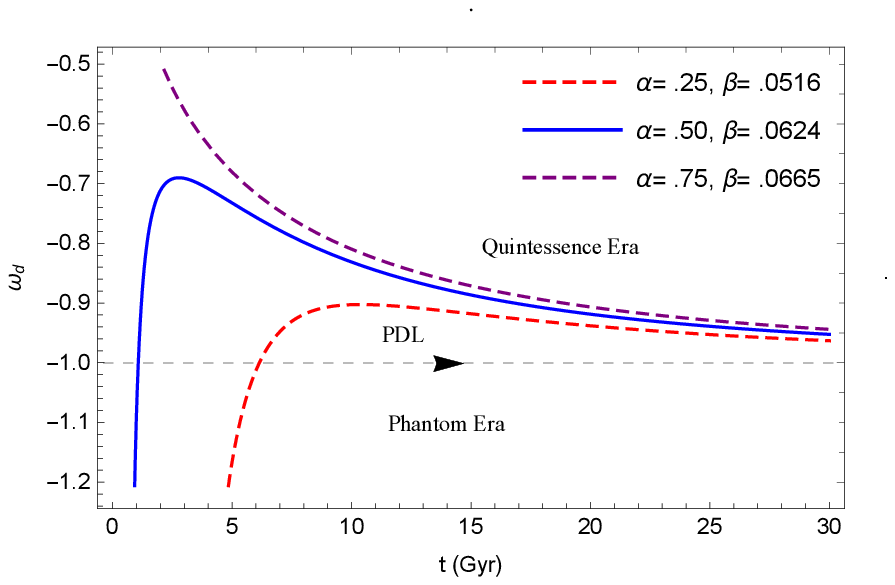}
	(b)\includegraphics[width=7cm,height=6cm,angle=0]{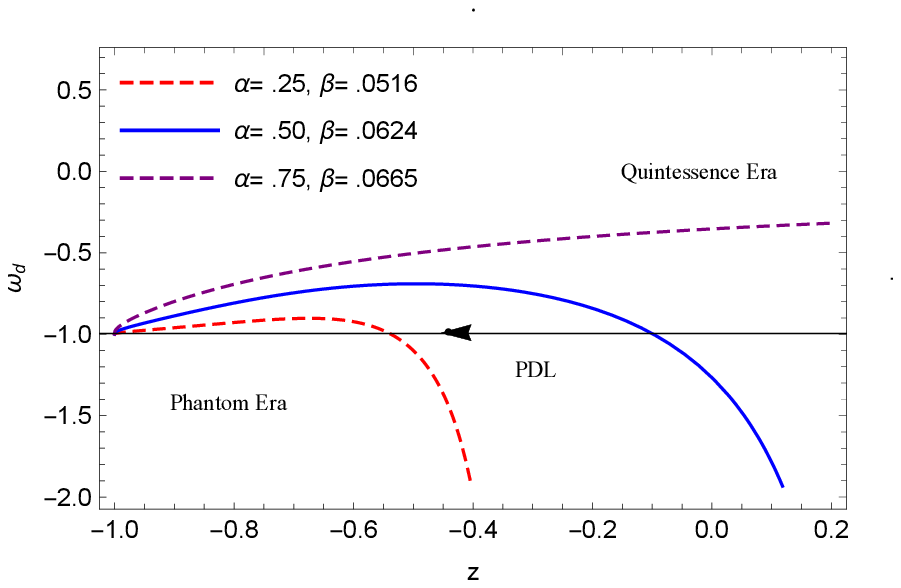}
	\caption{Variation of EoS parameter $\omega_{d} $ versus $ t $ and $z$}
	\label{fig1}.	
   \end{figure}
%%%%%%%%%%%%%%%%%%%%%%%%%%%%%%%%%%%%%%%%%%%%%%%%%%%%%%%%%%%%%%%%%%%%%%%%%%%%%%%%%%%%%%%%%%%%%%%%%%%%%%%%%%%%%%%%%%%%%%%%%%%%%%%%%
The redshift of a comoving cosmological source will change with time along the expansion of the universe. The monitoring of the redshift was 
first proposed by Sandage in 1962 \cite{ref54}, as a useful tool for testing of the cosmology. The expansion of universe can be directly 
measured by the redshift drift. Therefore, redshift is projected as a valuable addition to other DE probes \cite{ref55,ref56,ref57}. 
Now in our derived model, we use Hubble horizon as a IR cut-off, where $L=H^{-1}$. In general, the transformation between scale factor and redshift is given by $a=\frac{a_{0}}{(1+z)}$, where $ a_0 $ is the present value of $ a $. The redshift transformations are useful in validation of models with observational data. Here we obtained cosmological parameters in terms of $t$ as well as redshift in this paper.\\

The evolution of EoS parameter $\omega_{d}$ versus $t$ and $z$ for different values of $\alpha$ and $\beta$ are shown in fig.1(a)$\&$ 1(b).
The EoS parameter exists in both quintessence and phantom regions and also crosses the phantom divided line (PDL) ($\omega_{d}=-1$). Also, 
the latest cosmological data from SNIa (Supernova Legacy Survey, Gold Sample of Hubble Space Telescope) \cite {ref58,ref59}, 
CMB (WMAP, BOOMERANG) \cite {ref60} and large scale structure (SDSS) \cite {ref61} rule out that $\omega_{d}=-1$, they moderately support 
dynamically developing DE crossing the PDL (see \cite {ref62,ref63,ref64}). In cosmology, a redshift is an increase in the wavelength of electromagnetic radiation, with a corresponding fall in frequency and photon energy (such as light). Negative redshift, or blueshift, is the total opposite of positive redshift, with a reduction in wavelength and concomitant rise in frequency and energy. In our LRS Bianchi type-I model, $\omega_{d}=-1$, in late time as $z\to-1$ which suggests that matter in the Universe behaves like a perfect fluid initially. Later on the model is similar to a dark energy model and behaves like a quintessence model and finally approaches -1 then entering the phantom region. \\

As a consequence, our DE model agrees well with both well-established simulated predictions and current findings.The dynamical features of 
the model with HEL is most general and interesting feature as obtained from the power law or the exponential expansion. %Similarly, in fig 1b, our model predicts that the anisotropy will vanish over a sufficiently long period of time, and the cosmos will become isotropic.
Fig.1(b), shows that the EoS parameter approaches towards $\Lambda$CDM ($ \omega_{d}=-1$) in late time.
 
%%%%%%%%%%%%%%%%%%%%%%%%%%%%%%%%%%%%%%%%%%%%%%%%%%%%%%%%%%%%%% Figure 2 %%%%%%%%%%%%%%%%%%%%%%%%%%%%%%%%%%%%%%%%%%%%%%%%%%%%%%% 
 \begin{figure}[H]
	\centering
	(a)\includegraphics[width=7cm,height=7cm,angle=0]{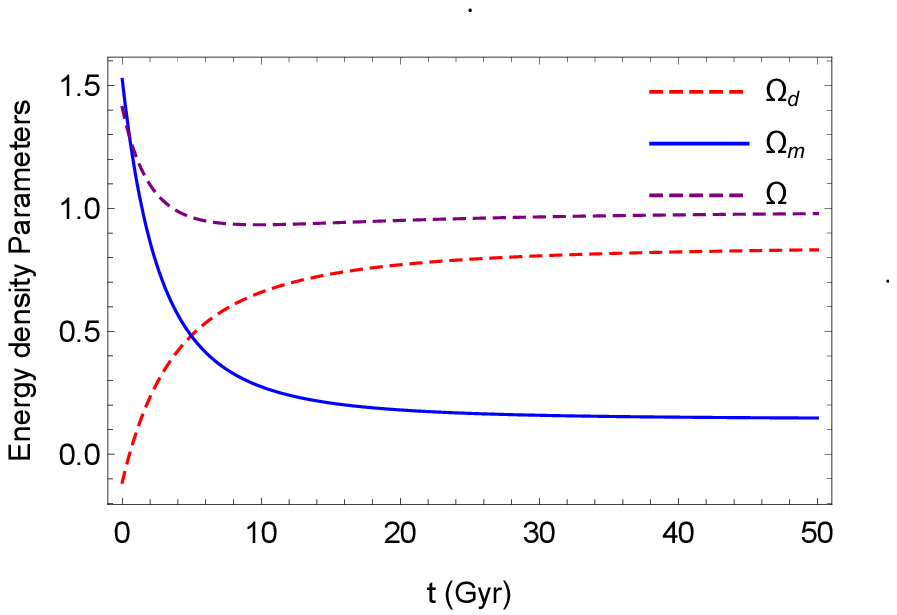}
	(b)\includegraphics[width=7cm,height=7cm,angle=0]{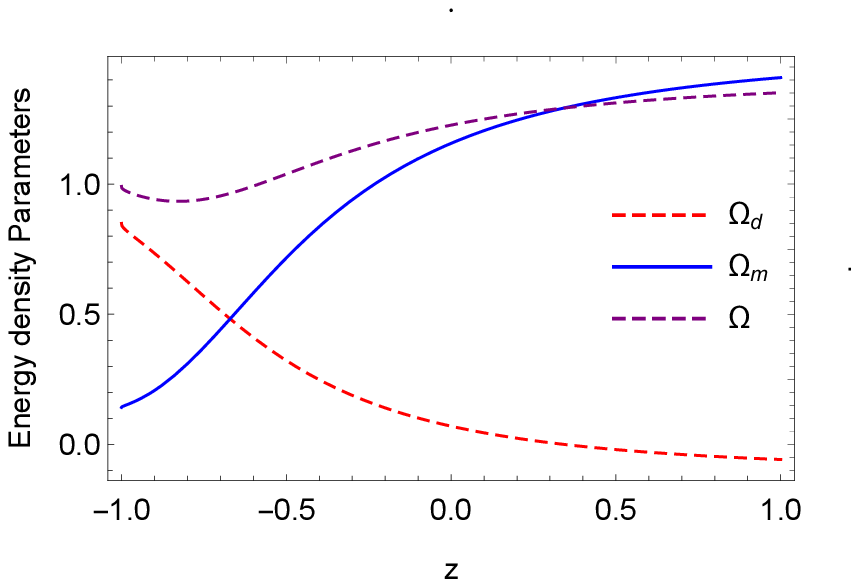}
	\caption{Variation of energy density parameters.}	`
	\label{fig2}	
\end{figure}
%%%%%%%%%%%%%%%%%%%%%%%%%%%%%%%%%%%%%%%%%%%%%%%%%%%%%%%%%%%%%%%%%%%%%%%%%%%%%%%%%%%%%%%%%%%%%%%%%%%%%%%%%%%%%%%%%%%%%%%%%%%%%%

Figure 2 describe the behaviour of energy density parameters with respect to time and redshift $z$ respectively. In the late time the total 
energy density  $\Omega\to1$ of the universe approaches to unity at low redshift region as clearly shown in Figure 2(a) $\&$ 2(b).

%%%%%%%%%%%%%%%%%%%%%%%%%%%%%%%%%%%%%%%%%%%%%%%%%%%%%%%%%%%%%%%% Figure 3 %%%%%%%%%%%%%%%%%%%%%%%%%%%%%%%%%%%%%%%%%%%%%%%%%%%
 
\begin{figure}[H]
	\centering
	(a)\includegraphics[width=7cm,height=7cm,angle=0]{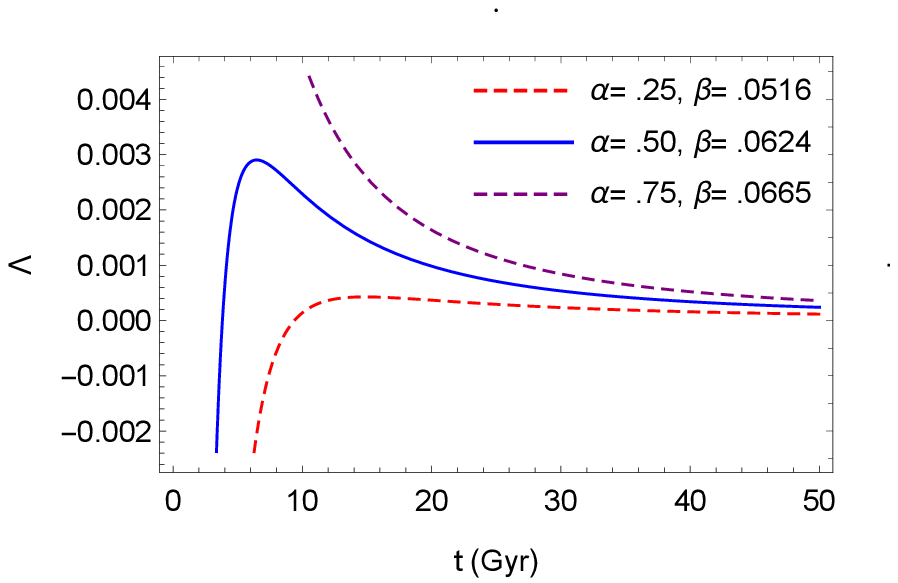}
	(b)\includegraphics[width=7cm,height=7cm,angle=0]{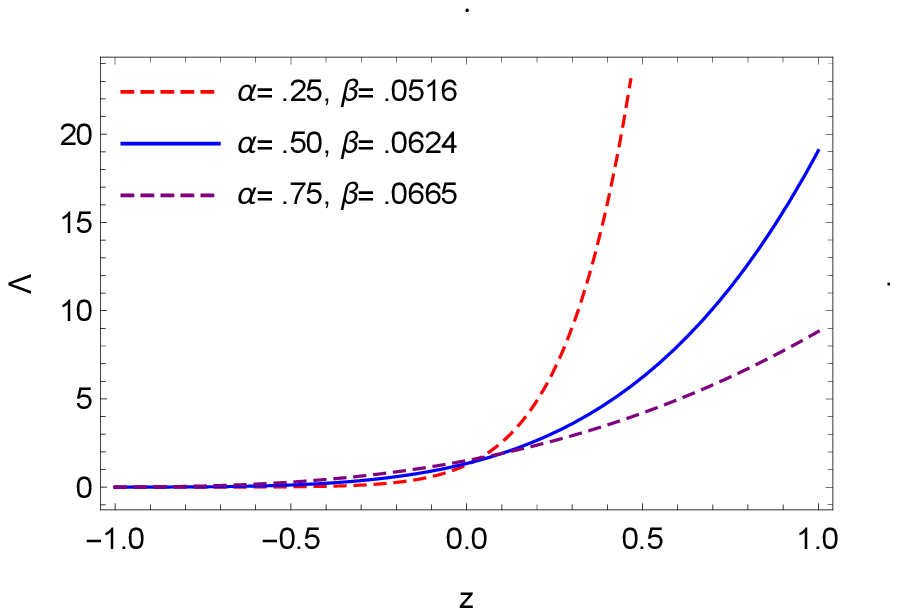}
	\caption{Variation of cosmological constant $(\Lambda) $.}
	\label{fig3}	
\end{figure}
%%%%%%%%%%%%%%%%%%%%%%%%%%%%%%%%%%%%%%%%%%%%%%%%%%%%%%%%%%%%%%%%%%%%%%%%%%%%%%%%%%%%%%%%%%%%%%%%%%%%%%%%%%%%%%%%%%%%%%%%%%%%%%%%%%%%%%%

Figures $3a$ $\&$ $3b$ depict the behavior of  the `cosmological constant $\Lambda$' versus $t$ and redshift $z$ respectively and show the 
universe's expanding mode. We observed that the cosmological parameter is the positive decreasing function of time and redshift $(z\to-1$) and 
approaches towards zero in late time. Which is a good agreement with recent SNe Ia observation \cite{ref65,ref66}. Authors \cite{ref67,ref68,ref69} 
propose the existence of a positive cosmological constant with a magnitude of $\Lambda (Gh/c3)\approx 10^{-123}$. The most impressive clarification 
for these info is that the dark energy compares to a positive cosmological constant.
These findings on the amplitude  and redshift of type Ia supernovas show that our universe is accelerating, with cosmological density produced 
by the cosmological expansion. The outcome  suggests a very minute positive value having magnitude  $10^{-123}$.

%%%%%%%%%%%%%%%%%%%%%%%%%%%%%%%%%%%%%%%%%%%%%%%%%%%%%%%%%%%%%%%%% Section 4 %%%%%%%%%%%%%%%%%%%%%%%%%%%%%%%%%%%%%%%%%%%%%%%%%%

\section{Statefinder}

The Statefinder is a geometrical diagnostic and allows us to characterize the properties of dark energy in a model independent manner. The Statefinder is dimensionless and is constructed from the scale factor of the Universe and its time derivatives only. In our model we demonstrate that the Statefinder diagnostic can effectively differentiate between different forms of dark energy. \\

The stability of DE models can  also be examined through the statefinder diagnostic pair $(r, s)$, which gives us an idea related to the dynamical 
nature of the model. The statefinder parameters $(r, s)$ \cite{ref70,ref71,ref72} are a pair of parameters that explore the dynamics of the 
universe's expansion by using the higher derivatives of the scale factor. In fact, the trajectories in the $(r, s) $ plane corresponding to 
different cosmological models show qualitative behaviour.

\begin{equation}\label{18}
r=\frac{\dddot{a}}{a H^3}=\frac{\alpha  (-3 \alpha -3 \beta  t+2)}{(\alpha +\beta  t)^3}+1
\end{equation}
\begin{equation}\label{19}
s=\frac{r-1}{3(-\frac{1}{2}+q)}=\frac{\alpha  (-3 \alpha -3 \beta  t+2)}{3 (\alpha +\beta  t)^3 \left(\frac{1}{2}-\frac{\alpha }
{(\alpha +\beta  t)^2}\right)}
\end{equation}
 
%The $(r ~s)$ trajectory for ($\alpha = 1, 1.5, 2.0$ and $\beta = 0.06699281858, 0.0621030872, 0.0522938602$) 

In the derived model, the ($r, s$) trajectories divide the region into two parts for the various value of the  $\alpha$ and $\beta$ as 
shown in figure 4a.  The region $r<1$ ,$s>0$ shows the quintessence era and $r>1$ ,$s<0$ shows the Chaplygin Gas (CG)region. The region 
$r=1$ ,$s=2/3$ describing the HDE and region $r=1$ ,$s=0$ approaches to $\Lambda$CDM model. In figure 4b, the trajectories $(r,q)$ show the
evolution of the universe. Our model starts with SCDM ($(r,q)=(1,0.5)$), $\Lambda$CDM  ($(r,q)=(1,0)$) and tends to SS ($(r,q)=(1,-1)$) model 
as shown in figure by the dots.

%%%%%%%%%%%%%%%%%%%%%%%%%%%%%%%%%%%%%%%%%%%%%%%%%%%%%%%%%%%%%%% Figure 4 %%%%%%%%%%%%%%%%%%%%%%%%%%%%%%%%%%%%%%%%%%%%%%%%%%%%%%%
\begin{figure}
	\centering
(a)\includegraphics[scale=1.0]{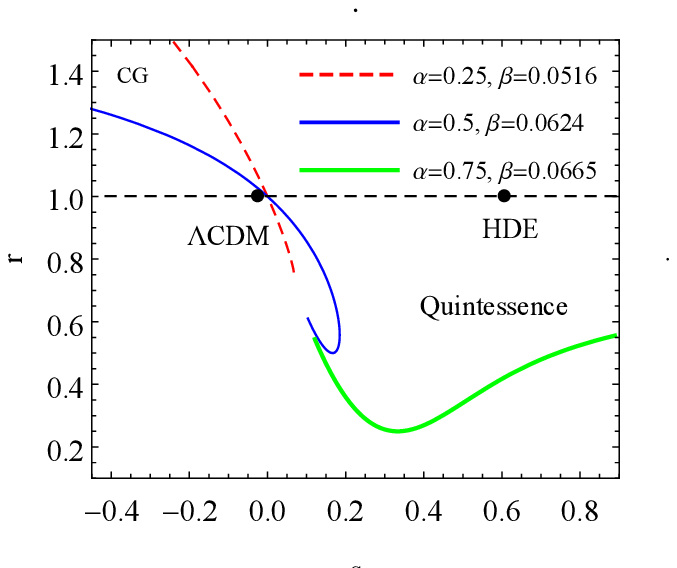}
(b)\includegraphics[scale=0.9]{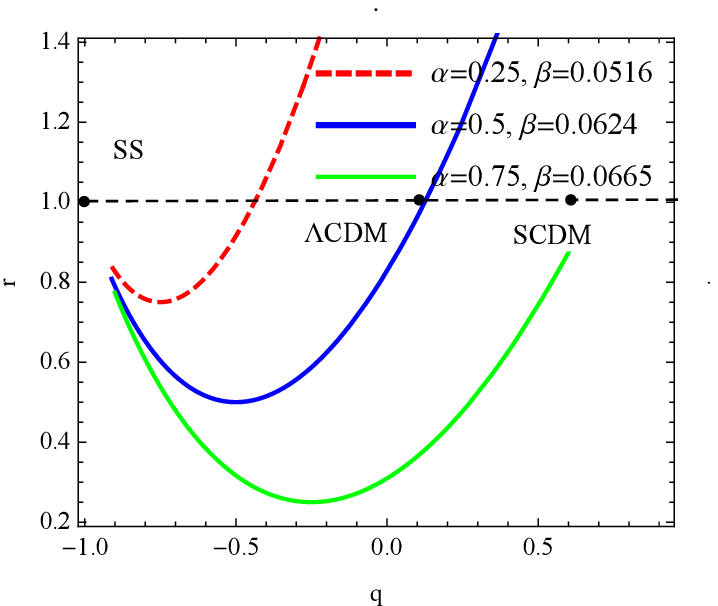}
	\caption{Plots of $ r $ versus $ s $ and $ r $ versus $ q $ . }
	\label{fig4}	
\end{figure}
%%%%%%%%%%%%%%%%%%%%%%%%%%%%%%%%%%%%%%%%%%%%%%%%%%%%%%%%%%%%%%%%%%%%%%%%%%%%%%%%%%%%%%%%%%%%%%%%%%%%%%%%%%%%%%%%%%%%%%%%%%%%%%%%%%%%

%%%%%%%%%%%%%%%%%%%%%%%%%%%%%%%%%%%%%%%%%%%%%%%%%%%%% Section 5 %%%%%%%%%%%%%%%%%%%%%%%%%%%%%%%%%%%%%%%%%%%%%%%%%%%%%%%%%%%%%%%
\section{$\omega_{d}-\omega_{d}^{'}$  plane}

The EoS parameter of dark energy $\omega_{d} = p_{d}/\rho_{d}$, plays a key role in observational cosmology. According to the distinct 
region in the $\omega_{d} -\omega'_{d}$ plane, 
Caldwell and Linder \cite{ref73, ref74} categorized the quintessence models defined by the quantities $\omega_{d}$ and  $\omega'_{d}$ and 
divide into two parts "thawing" and "freezing," region. Some researchers have recently examined  the evolution of quintessence DE models 
on the $\omega_{d}-\omega'_{d}$ plane, where $\omega'_{d}$ is the time derivative of $\omega_{d}$ with respect to $\log a$. 

\begin{equation} \label{20}
\omega'_{d}=\frac{d \omega_{d}}{d \log a}=\frac{\alpha  t \left(3 \alpha ^2 k^2-3 k m (\alpha -\beta  t) 
	(\alpha +\beta  t)-2 \beta  m^2 (\alpha +\beta  t)^2\right)}{3 (\alpha +\beta  t)^3 (\alpha  k-m (\alpha +\beta  t))^2}
\end{equation}
 
In our derived model the quintessence region asymptotically approaches to zero.
Later this analysis was extended by many researchers to other dynamic DE models.
Clearly, a fixed point at (-1,0) is the $\Lambda$ CDM in the $\omega_{d}-\omega'_{d}$ plane.

Finally we analyzed  the evolutionary behavior of the EoS parameter of the HDE in figure $5$. The trajectories lies in both freezing and 
thawing regions for three values of parameters $\alpha$ and $\beta$ as clearly shown in figure.

%%%%%%%%%%%%%%%%%%%%%%%%%%%%%%%%%%%%%%%%%%%%%%%%%%%%%%%% Figure 5 %%%%%%%%%%%%%%%%%%%%%%%%%%%%%%%%%%%%%%%%%%%%%%%%%%%%%%%%%%%%
\begin{figure}
	\centering
	\includegraphics[scale=0.9]{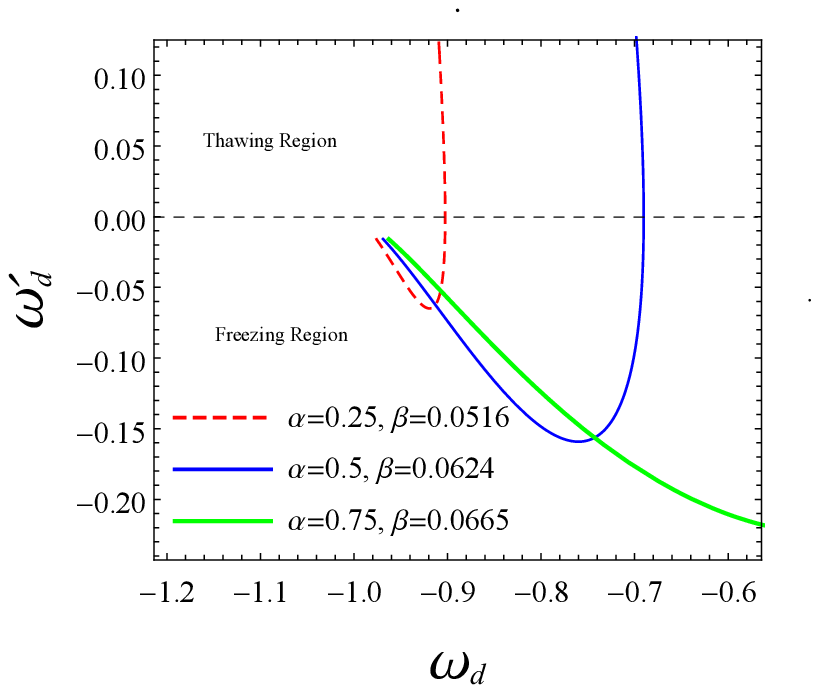}
	\caption{Plot of $ \omega_{d}-\omega'_{d} $. }
	\label{fig5}	
\end{figure}
%%%%%%%%%%%%%%%%%%%%%%%%%%%%%%%%%%%%%%%%%%%%%%%%%%%%%%%%%%%%%%%%%%%%%%%%%%%%%%%%%%%%%%%%%%%%%%%%%%%%%%%%%%%%%%%%%%%%%%%%%%%%%%%%%%%%%%%%%%%

%%%%%%%%%%%%%%%%%%%%%%%%%%%%% Section 6 %%%%%%%%%%%%%%%%%%%%%%%%%%%%%%%%%%%%%%%%%%%%%%%%%%%%%%%%%%%%%%%%%%%%%%%%%%%%%%
\section{Scalar Field DE Models}

%%%%%%%%%%%%%%%%%%%%%%%%%%%%%%%%%%%%%%%%%%%%%%%%% Subsection 6.1 %%%%%%%%%%%%%%%%%%%%%%%%%%%%%%%%%%%%%%%%%%%%%%%%%%%%%%%%%%

\subsection{Tachyon scalar field model of HDE }

To depicts the accelerated expansion and evaluation of universe, the tachyon field (TF) is one of the DE candidates \cite{ref75,ref76}. 
Some tachyon condensate by an powerful Lagrangian density \cite{ref77,ref78} recommends a tachyon field is given by
\begin{equation}\label{21}
L=-V(\phi)\sqrt{1+\partial_{i} \phi \ \partial^{i} \phi}
\end{equation}
Tachyon is observed the most important DE constituent which explain the accelerated expansion of the universe. The EoS parameter of the 
tachyon DE matter distribution lies between $-1$ and $0$  \cite{ref79}. The `energy density $\rho_{d} $' and `pressure $ p_{d} $' associated 
with Scalar field ($\phi $) and `scalar potential $ V(\phi)$' for tachyon matter distribution are read as:
\begin{equation}\label{22}
p_{d}= \left(1-\dot{\phi}^2\right)^{1/2} V(\phi)
\end{equation}
\begin{equation}\label{23}
\rho_{d}=\frac{V(\phi)}{\sqrt{1-\dot{\phi}^2}} 
\end{equation}
 Here, $ \dot{\phi}^2$  is the scalar field and $ V(\phi)$  is the potential respectively.\\
From Eqs. (22) - (23) and using Eqs. (11) and (13), the scalar $\phi$ and potential $V(\phi)$ are determined  as:
\begin{equation}\label{24}
\phi= \int \sqrt{\frac{2}{3}} \sqrt{-\frac{\alpha  (m-k (\alpha +\beta  t))}{(\alpha +\beta  t) \left(k (\alpha +\beta  t)^2-
\alpha  m\right)}} \, dt+c_{1}
\end{equation}
\begin{equation}\label{25}
V(\phi)=\sqrt{3} \sqrt{\frac{\left(k (\alpha +\beta  t)^2-\alpha  m\right) \left(-2 \alpha  k+3 \left(k (\alpha +\beta  t)^2-
\alpha  m\right)+\frac{2 \alpha  m}{\alpha +\beta  t}\right)}{t^4}}
\end{equation}

%%%%%%%%%%%%%%%%%%%%%%%%%%%%% Figure 6 %%%%%%%%%%%%%%%%%%%%%%%%%%%%%%%%%%%%%%%%
\begin{figure}[H]
	\centering
	(a)\includegraphics[scale=0.7]{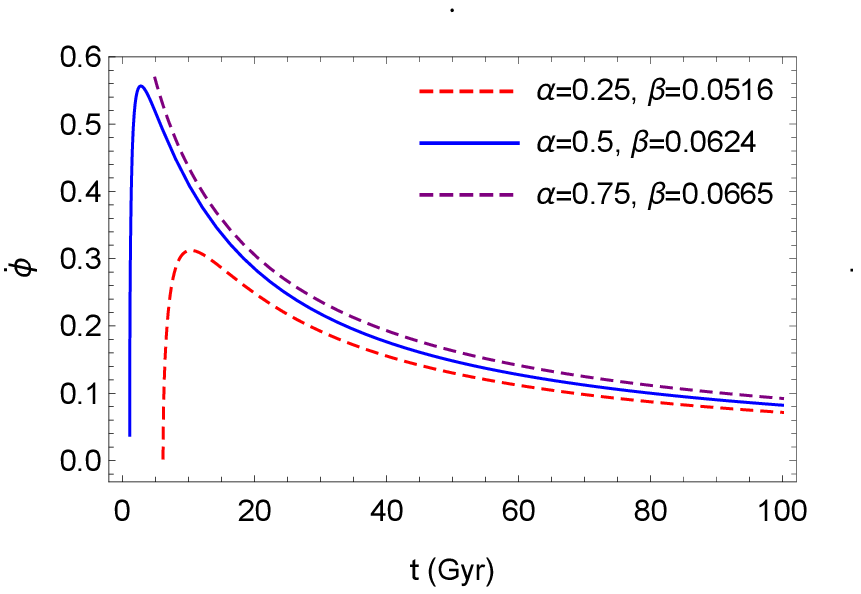}
	(b)\includegraphics[scale=0.7]{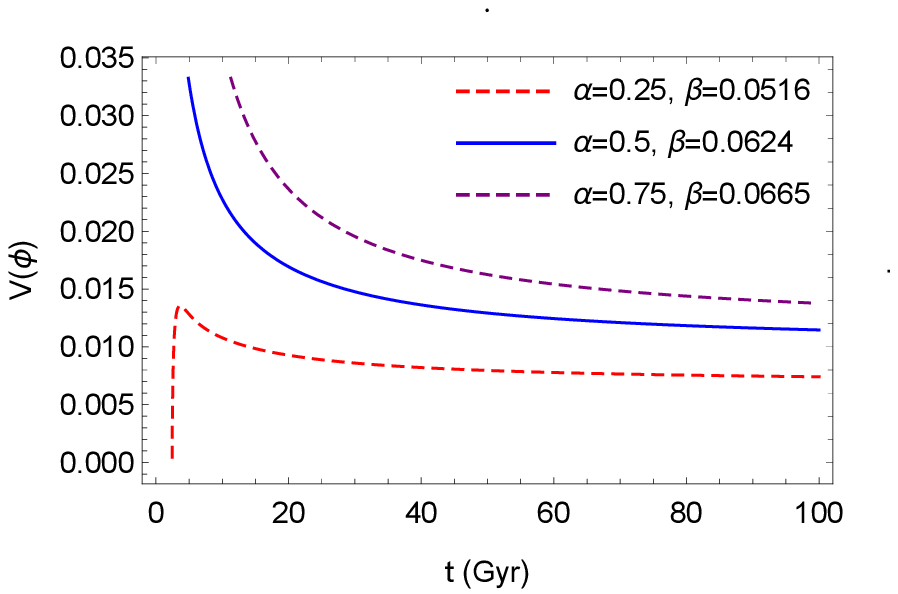}
	\caption{Plots of $ \dot{\phi} $ and  $ V(\phi) $ for tachyon field . }
	\label{fig6}	
\end{figure}

%%%%%%%%%%%%%%%%%%%%%%%%%%%%%%%%%%%%%%%%%%%%%%%%%%%%%%%%%%%%%%%%%%%%%%%%%%%%%%%%

 We likewise present an extensive examination of the cosmological results of homogeneous tachyon matter coinciding with non-relativistic 
 matter and radiation, with an emphasis on the converse square and exponential potentials for the tachyonic scalar field. Figures $6a$ and $6b$ 
 show that the scalar field $\phi$ and  potential $V(\phi)$ both remains positive as time advances in our determined model.
The dynamical behaviour of the tachyon field associated with an exponential potential is presented, and we demonstrate that the pre-inflation 
dynamics in modified loop cosmology and standard loop quantum cosmology are remarkably similar. 

%%%%%%%%%%%%%%%%%%%%%%%%%%%%%%%%%%%%%%%%%%%%%%%%%%%%%%%%%% Subsection 6.2 %%%%%%%%%%%%%%%%%%%%%%%%%%%%%%%%%%%%%%%%%%%%%%%%%%%%%%%%%%%

\subsection{Quintessence scalar field model of HDE }
Quintessence is a conceptual formulation of DE characterized by homogeneous variable `Scalar field and the  potential', which are essential 
for the expansion of the universe \cite{ref80}. For the quintessence scalar $\phi $, the general formulation of action is as follows:
\begin{equation}\label{26}
S = \int{d^{4}x \sqrt{-g}\left(-V(\phi)-\frac{1}{2} g^{ij} \partial_{i} \phi \partial_{j} \phi\right)}
\end{equation}

The EoS parameter in the quintessence model indicates that the accelerated expansion of the universe exists for $-1\leq\omega \leq -\frac{1}{3}$ 
\cite{ref77}. For the quintessence model, the relations of $\rho_{d} $ and $ p_{d} $ in the form of `$ \phi $' and `$ V(\phi) $' are established 
as \cite{ref63}:

\begin{equation}\label{27}
p_{d}=\frac{1}{2}\left(\dot{\phi}^2-2V(\phi)\right)
\end{equation}  
\begin{equation}\label{28}
\rho_{d}=\frac{1}{2}\left(\dot{\phi}^2+2V(\phi)\right)
\end{equation}
From Eqs. (11), (13), (27) and (28), the scalar $\phi$ and potential $V(\phi)$ for the `quintessence model' are obtained as
\begin{equation}\label{29}
\phi=\int \sqrt{\frac{3 \left(-\frac{2 \alpha  (m-k (\alpha +\beta  t))}{3 (\alpha +\beta  t) \left(k (\alpha +\beta  t)^2-
\alpha  m\right)}-1\right) \left(k (\alpha +\beta  t)^2-\alpha  m\right)}{t^2}+\frac{3 \left(k (\alpha +\beta  t)^2-\alpha  m\right)}{t^2}} \, dt+c_{2}
\end{equation}\label{30}  
\begin{equation}
V(\phi)=\frac{k \left(3 \alpha ^3+3 \beta ^3 t^3+\alpha ^2 (9 \beta  t-1)+\alpha  \beta  t (9 \beta  t-1)\right)+
\alpha  m (-3 \alpha -3 \beta  t+1)}{t^2 (\alpha +\beta  t)}
\end{equation}

%%%%%%%%%%%%%%%%%%%%%%%%%%%%% Figure 7 %%%%%%%%%%%%%%%%%%%%%%%%%%%%%%%%%%%%%%%%
\begin{figure}[H]
	\centering
	(a)\includegraphics[scale=0.7]{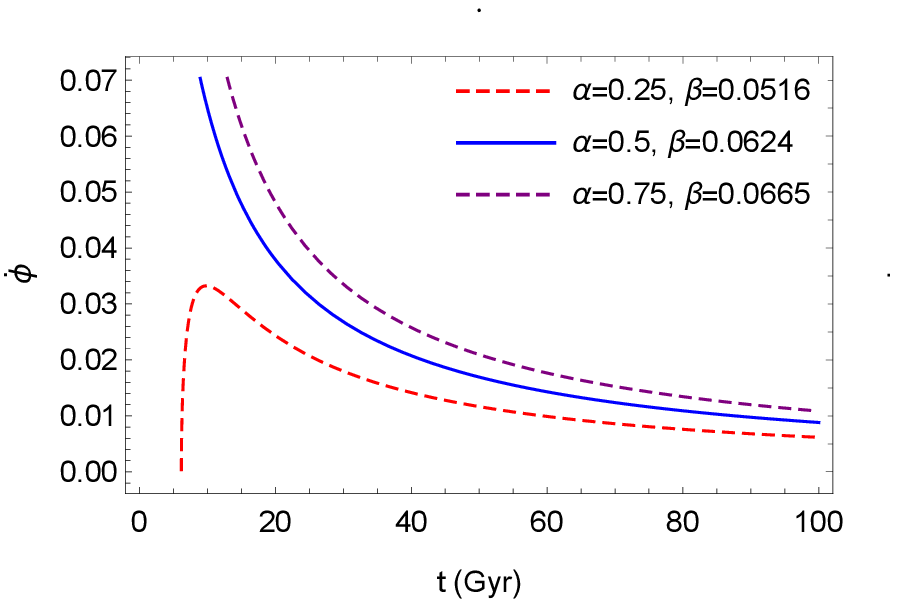}
	(b)\includegraphics[scale=0.7]{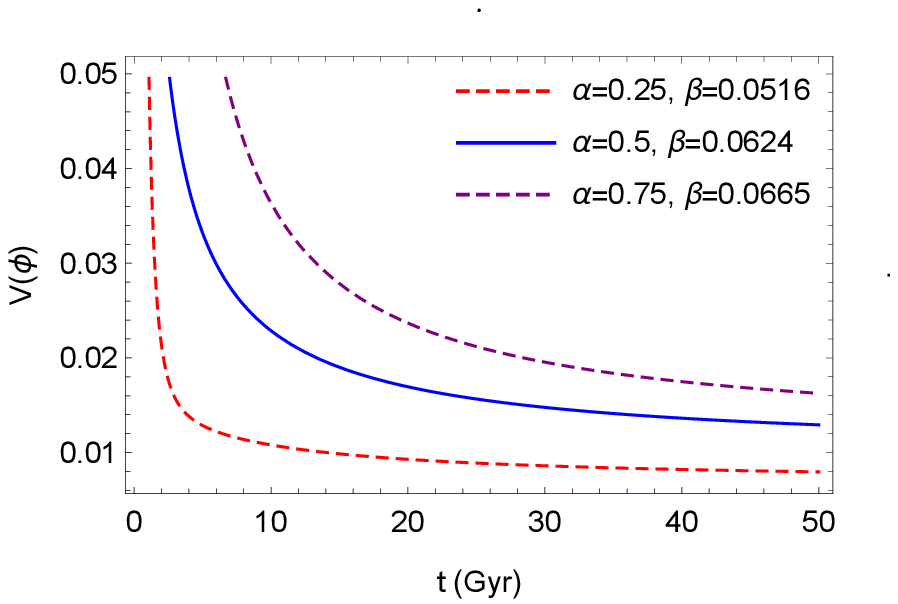}
	\caption{Plots of $ \dot{\phi} $ and  $ V(\phi) $ for quintessence field . }
	\label{fig7}	
\end{figure}

%%%%%%%%%%%%%%%%%%%%%%%%%%%%%%%%%%%%%%%%%%%%%%%%%%%%%%%%%%%%%%%%%%%%%%%%%%%%%%%%

Figures $7a$ and $7b$ depict the growth of the `scalar field $\phi$' and ` potential $ V(\phi) $' with time $t$ for all three values of $\alpha$ 
and $\beta$ in our derived model. The Scale field and  potential diminish with time and tends to be low positive at the present epoch. 

%%%%%%%%%%%%%%%%%%%%%%%%%%%%%%%%%%%%%%%%%%%%%%%%%%%%%% Section 7 %%%%%%%%%%%%%%%%%%%%%%%%%%%%%%%%%%%%%%%%%%%%%%%%%%%%%%%% 

\section{Deceleration parameter for the Model}

From the Table-1, we observe that $\alpha$ increases with the increase of $\beta$. For $0 < \alpha \leq 1$, $\beta$ increases  
but for $\alpha > 1$, $\beta$ starts decreasing.  
It is worth mentioned that for $0<\alpha<1$, our models 
show the transitioning scenario whereas for $\alpha \geq 1$, the models are in accelerating phase only. \\

Figure $8(a)$ depicts the variation of DP ($q$) with time $t$. Because DP ($q$) lies between the range $-1 < q < 0$, SNe Ia data 
reveals the expansion of the present universe in accelerating mode. As a result, the value of the decelerating parameter matches recent data. \\

We have demonstrated in Fig. $8(b)$ that expansion of $q(z)$ takes place with the change of signature flipping from a decelerating to an 
accelerating phase. We also observed that $q$ tends to $-1$ as $z$ approaches to $-1$. SNe Ia data shows the progression from previous deceleration 
to current acceleration. The transition point $z_{t} = 0.46 \pm 0.13$ at $(1\;\sigma)$ c.l. was investigated in 2004 by the HZSNS group \cite{ref68}.
In 2007, this value of $z_{t}$ was improved by $z_{t} = 0.43 \pm 0.07$ at $(1\;\sigma)$ c.l. \cite{ref82}. SNLS first year data set \cite{ref83} 
combined with ESSENCE supernova data \cite{ref84}, provide a progress redshift $z_{t} \sim 0.6 (1\; \sigma)$ in improved agreement with the 
flat $\Lambda$CDM model ($z_{t} = (2\Omega_{d}/\Omega_{m})^{\frac{1}{3}} 
- 1 \sim 0.66$).\\

Again $q(z)$ is reconstructed by the joined $(SNIa + CC + H_{0})$, and obtained the transition redshift 
$z_{t} = {0.69}^{+0.09}_{-0.06}, {0.65}^{+0.10}_{-0.07}$ and ${0.61}^{+0.12}_{-0.08}$ within $(1 \sigma)$ \cite{ref85} and found 
consistent with past outcomes \cite{ref86,ref87,ref88,ref89,ref90} including the $\Lambda CDM$ expectation 
$z_{t} \approx 0.7$. $0.6 \leq z_{t} \leq 1.18$ ($2\sigma$, joint examination ) \cite{ref91} is the other limit of transition redshift.\\

From the figure $8(b)$, the transitioning redshifts occur at 
$z_{t} \cong 0.6921$ (for $\alpha = 0.25$ \& $\beta = 0.9516$);  $z_{t} \cong 0.4443$ (for $\alpha = 0.50$ \& $\beta = 0.0624$), and 
$z_{t} \cong 0.995$ (for $\alpha = 0.75$ \& $\beta = 0.0665$) for above three cases which are found to be well consistent 
with the recent 36 observational Hubble data (OHD) provides the redshift range $0.07 \leq z \leq 2.36$ \cite{ref92}.
Comprising 740 SNIa with JLA indicates the redshift range `$0.01 \leq z \leq 1.30$'. 
Obviously, our results are in good accordance with those reported in Refs. \cite{ref68,ref84,ref91,ref92}. \\
%%%%%%%%%%%%%%%%%%%%%%%%%%%%%%%%%%%%%%%%%%%%%%%%%%%%%%%%%%%% Figure 8 %%%%%%%%%%%%%%%%%%%%%%%%%%%%%%%%%%%%%%%%
\begin{figure}[H]
	\centering
	(a)\includegraphics[scale=0.80]{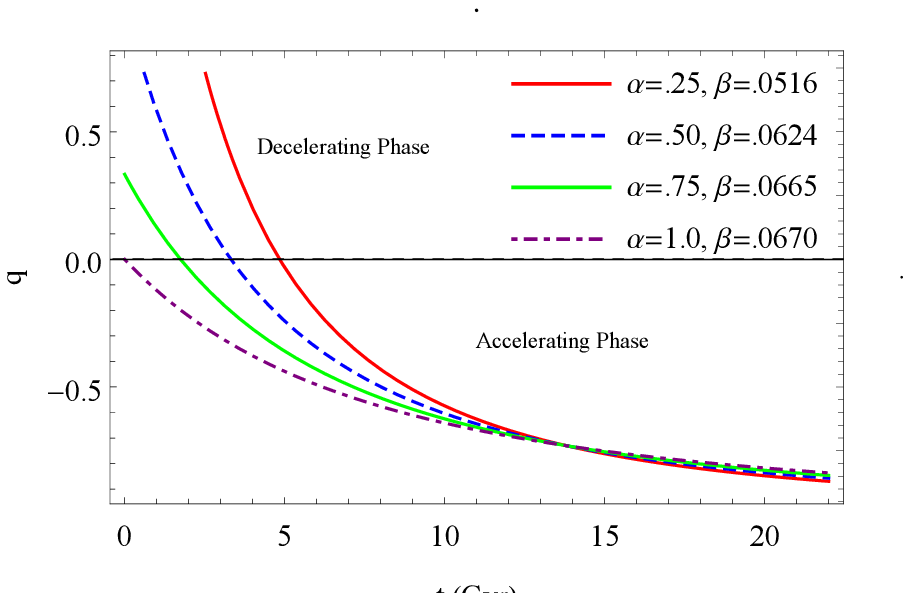}
	(b)\includegraphics[scale=0.80]{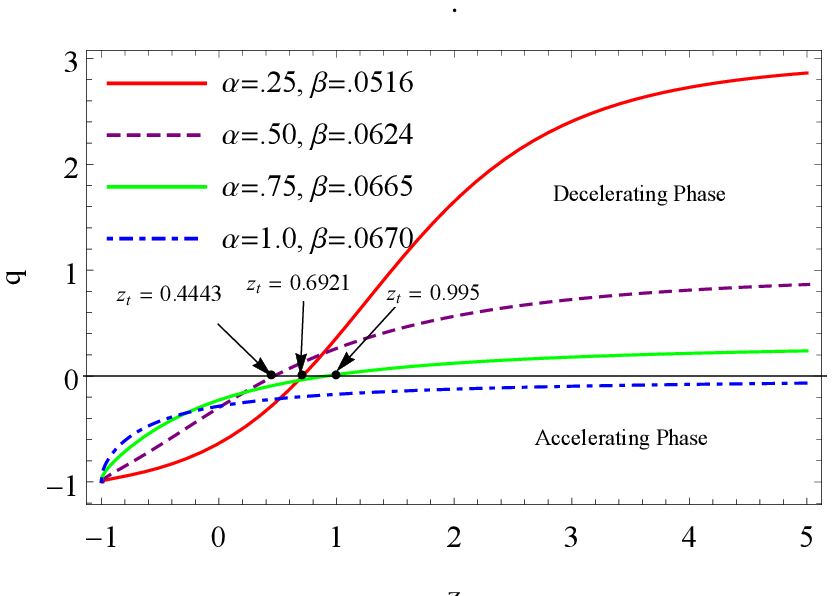}
	\caption{Plots of Deceleration parameter $q $ versus  $ t $ and redshift $ z $. }
	\label{fig8}	
\end{figure}
%%%%%%%%%%%%%%%%%%%%%%%%%%%%%%%%%%%%%%%%%%%%%%%%%%%%%%%%%%%%%%%%%%%%%%%%%%%%%%%%%%%%%%%%%%%%%%%%%%%%%%%%%%%%%%%%%%

%%%%%%%%%%%%%%%%%%%%%%%%%%%%%%%%% Table-1 %%%%%%%%%%%%%%%%%%%%%%%%%%%%%%%%%%%%%%
	\begin{table}[H]
	\caption{\small The behaviour of physical parameter}
	\begin{center}
		\begin{tabular}{|c|c|c|c|c|c|c|c|c|c|c| }
			\hline
		 $ \alpha $  & 	 $ \beta $  & 	 $ q $ & 	 $ \omega $  & 	 $ \Omega _d $ & 	 $ \Omega _m $ & 	  $ \Omega $  &  $ r $  & 	$ s $  & 	 $ \omega _{d}^{'} $   \\
		\hline
    \tiny 0.25 &  \tiny 0.0516124 &  \tiny 1.00372 &  \tiny 4.93611 &  \tiny -0.115795 &  \tiny 2.06244 &  \tiny 1.94665 &  \tiny 6.33415 &  \tiny 3.52981 &  \tiny 59.0776 \\
     \hline
	 \tiny 0.5 &  \tiny 0.0623788 &  \tiny 0.280993 &  \tiny -0.711735 &  \tiny 0.232561 &  \tiny 0.863276 &  \tiny 1.09584 &  \tiny 1.25779 &  \tiny -0.392361 &  \tiny 0.232003 \\
	 \hline
      \tiny 0.75 &  \tiny 0.0664251 &  \tiny -0.0377516 &  \tiny -0.494967 &  \tiny 0.386196 &  \tiny 0.539123 &  \tiny 0.925319 &  \tiny 0.293123 &  \tiny 0.438168 &  \tiny -0.224348 \\
     \hline
      \tiny 1. &  \tiny 0.0669928 &  \tiny -0.222349 &  \tiny -0.53636 &  \tiny 0.475172 &  \tiny 0.400258 &  \tiny 0.87543 &  \tiny 0.0385825 &  \tiny 0.443653 &  \tiny -0.124511 \\
      \hline
       \tiny 1.25 &  \tiny 0.0653375 &  \tiny -0.344267 &  \tiny -0.589897 &  \tiny 0.533937 &  \tiny 0.326571 &  \tiny 0.860507 &  \tiny -0.0173261 &  \tiny 0.401661 &  \tiny -0.0699145 \\
      \hline
       \tiny 1.5 &  \tiny 0.0621031 &  \tiny -0.431397 &  \tiny -0.635868 &  \tiny 0.575933 &  \tiny 0.282205 &  \tiny 0.858139 &  \tiny -0.00564742 &  \tiny 0.359906 &  \tiny -0.041409 \\
      \hline
       \tiny 1.75 &  \tiny 0.0576721 &  \tiny -0.497056 &  \tiny -0.673521 &  \tiny 0.607581 &  \tiny 0.253154 &  \tiny 0.860735 &  \tiny 0.0304176 &  \tiny 0.324149 &  \tiny -0.0255545 \\
      \hline
      \tiny 2. &  \tiny 0.0522939 &  \tiny -0.54846 &  \tiny -0.704391 &  \tiny 0.632358 &  \tiny 0.232952 &  \tiny 0.86531 &  \tiny 0.0744813 &  \tiny 0.294247 &  \tiny -0.0161907 \\
      \hline
       \tiny 2.25 &  \tiny 0.0461414 &  \tiny -0.589887 &  \tiny -0.729989 &  \tiny 0.652325 &  \tiny 0.218254 &  \tiny 0.87058 &  \tiny 0.119843 &  \tiny 0.269189 &  \tiny -0.0103853 \\
      \hline
       \tiny 2.5 &  \tiny 0.0393408 &  \tiny -0.624037 &  \tiny -0.751496 &  \tiny 0.668786 &  \tiny 0.207177 &  \tiny 0.875963 &  \tiny 0.163705 &  \tiny 0.248003 &  \tiny -0.00664547 \\
      \hline
       \tiny 2.75 &  \tiny 0.0319872 &  \tiny -0.65271 &  \tiny -0.769794 &  \tiny 0.682606 &  \tiny 0.198589 &  \tiny 0.881195 &  \tiny 0.204962 &  \tiny 0.229904 &  \tiny -0.00416254 \\
      \hline
       \tiny 3. &  \tiny 0.0241546 &  \tiny -0.677148 &  \tiny -0.785542 &  \tiny 0.694385 &  \tiny 0.191773 &  \tiny 0.886159 &  \tiny 0.243268 &  \tiny 0.214284 &  \tiny-0.00247481 \\
	 \hline												
				
		\end{tabular}
	\end{center}
\end{table}

%%%%%%%%%%%%%%%%%%%%%%%%%%%%%%%%%%%%%%%%%%%%%%%%%%%%%%%%%%%%%%%%%%%%%%%%%%%%%%%%
%%%%%%%%%%%%%%%%%%%%%%%%%%%%%%%% Section 8 %%%%%%%%%%%%%%%%%%%%%%%%%%%%%%%%%%%
\section{Conclusion}
In this paper we have developed an anisotropic dark energy cosmological model in the framework of the LRS Bianchi type-I metric. The anisotropic 
behaviour of the model is simulated by taking HSF scale factor. In this model we are able to get the exact solutions to the field equations by 
using the hybrid scale factor. This manuscript depends on the HP and dimensional analysis and explored various parameters in the context of HDE, 
like EoS parameter, density parameter, and cosmological constant. In addition, the state finder trajectories and $\omega_{d}-\omega_{d}{'}$ 
plane are also diagnosed. We also analyzed the Hubble parameter for HDE, for both,  theoretical and  observational aspects.\\
 
The main findings of our model are as follows.\\
\begin{itemize}

\item  
The EoS parameter, which is a function of time and redshift, has been displayed in Figs. $1(a)$ and $1(b)$. Now in our derived model, 
we have obtained a quintessence and phantom region  and crosses the Phantom 
divided line for three values of ($\alpha$, $\beta$) are shown in Figs. $1(a)$ and $ 1(b)$. From the graph it is clear that the universe is under 
the influence of dark energy as the equation of state predicted accelerated expansion phase  

\item  
Figs. $2a$ $\&$ $2b$ depict the overall density parameter, which has been matched with $\Lambda CDM$ and approaches to Unity i.e. $\Omega\equiv1$ for the 
different values of ($\alpha$, $\beta$).\\ 

\item  
Figures $3a$ $\&$ $3b$ depict the evolution of the cosmological constant versus $t$ and $z$ respectively. Here $\Lambda$ is a positive decreasing 
function and approaches towards zero.	

\item  As illustrated in Figs. $4a$ and $4b$, the trajectories of ` $(r,s)$ and $(r,q)$' for various values of $\alpha$ and $\beta$ have been displayed. 
During the early time, the statefinder trajectory $(r,s)$ plane is separated into two regions: `` $r < 1, s > 0$ quintessence region and 
the $r > 1, s < 0$ Chaplygin Gas", and its approaches to $\Lambda$CDM $(r, s) = (1, 0)$ in the late time. The trajectory $(r, q)$ evolves from the region of (SCDM) in 
the $(r-q)$ plane and approaches the SS model in late time.

\item  In Fig. $5$ we have analyzed the evolutionary behaviour of the EoS parameter of the HDE models by constructing the  
$\omega_{d}-\omega_{d}^{'}$ plane and found that $\omega_{d}^{'}<0$ for $\omega_{d}<0$  for the HDE model implying the plane to lie in 
the freezing region as well as  $\omega_{d}^{'}>0$ for $\omega_{d}<0$ thawing region.

\item  We have also constructed the potential as well as dynamics of the quintessence and tachyon scalar field see Figs $6$ \& $7$.

\item Figure $8(b)$ shows the transition redshift from decelerated to accelerated expansion occurs at 
$z_{t} \cong 0.6921$ (for $\alpha = 0.25$ \& $\beta = 0.9516$);  $z_{t} \cong 0.4443$ (for $\alpha = 0.50$ \& $\beta = 0.0624$), and 
$z_{t} \cong 0.995$ (for $\alpha = 0.75$ \& $\beta = 0.0665$) for above three cases which are in good agreement with those reported in 
Refs. \cite{ref68,ref84,ref91,ref92}.

\end{itemize} 

As a result, our findings support recent observational data that revealed the current isotropic state. This demonstrates that our model 
closely matches current observations.
%%%%%%%%%%%%%%%%%%%%%%%%%%%%%%%%%%%%%%%%%%%%%%%%%%%%%%%%%%%%%%%%%%%%%%%%%%%%%%%%%%%%%%%%%%%%%%%%%%%%%%%%%%%%%
\section*{Credit authorship contribution statement}
Anirudh Pradhan: Ideas, Formulation and writing- original draft preparation. Vinod Kumar Bhardwaj: Application of mathematical formulation, 
computation \& drawing figures. Archana Dixit: Writing- reviewing, validation \& editing. Symala Krishannair: Final draft preparation.

\section*{Declaration of competing interest}
The authors declare that they have no known competing financial interests or personal relationships that could have appeared to influence the work reported in this paper.
   
   %%%%%%%%%%%%%%%%%%%%%%%%%%%%%%%%%%%%%%%%%%%%%%%%%%%%%%%%%%%%%%%%%%%%%%%%%%%%%%%%%%%%%%%%%%%%%%%%%%%%%%%%%%%%%%%
\section*{Acknowledgments}
The author (AP) thanks the IUCAA, Pune, India for providing the facility under visiting associateship. {\bf We are so grateful to the reviewers for their many valuable suggestions and comments that significantly improved the paper.}
%%%%%%%%%%%%%%%%%%%%%%%%%%%%%%%%%%%%%%%%%%%%%%%%%%%%%%%%%%%%%%%%%%%%%%%%%%%%%%%%%%%%%%%%%%%%%%%%%%%%%%%%%%%%%%%%  

%%%%%%%%%%%%%%%%%%%%%%%%%%%%%%%%%%%%%%%%%%%%%%%%%%%%%%%%%%%%%%%%%%%%%%%%%%%%%%%%%%%%%%%%%%%%%%%%%%%%%%%%%%%%%%%%%%%%%%%%%%%%%%%%%%%%%%%%%%%% 

 \end{document}